\newcommand{\beq}{\begin{equation}}
\newcommand{\eq}{\end{equation}}
\newcommand{\pl}{\partial}
\documentstyle[dinA4,12pt]{article}

\begin{document}

\makebox[13.5cm][r]{LMU--TPW 92--27}
\vspace{3cm}
\Large

\begin{center}
{\bf A DEFORMED POINCAR\'E ALGEBRA \\ ON A CUBIC LATTICE}}

\vspace{2cm}

{\large {\it Sebastian Sachse\\ \it Sektion Physik der Universit\"at M\"unchen
\\
\it Lehrstuhl Professor Wess, Theresienstra\ss e 37\\
\it D-8000 M\"unchen 2, Germany }}

\vspace{2cm}
{\large \today}
\end{center}
\vspace{4cm}

\begin{abstract}
Replacing the continuous space by a cubic lattice
we find a deformation of the Poincar\'e algebra.
A deformation of the relativistic mass operator is
shown to be a Casimir of the algebra. The real structure is preserved.
\end{abstract}
\newpage
\normalsize

\parskip3ex

Recently deformations of the Poincar\'e algebra have been under
intense investigation \cite{schl,ogi,luk}.
Our approach to this problem has been
inspired by a work of Bonechi et al. on the linear chain \cite{bon}, who have
used a relation between certain algebra deformations and lattice structure.
Other aspects of this connection have been given in \cite{dim}.

We start with the relativistic covariant Klein--Gordon equation.
In order to find the analogous equation on a lattice, we replace
differentiations by difference quotients.
We apply this to three space coordinates
but leave the time axis continuous.
With a constant spacing $a$ we get for the Klein--Gordon equation
\begin{eqnarray}
\ddot{\phi}_{\vec{n}} & = & \omega^2 \,\sum_{i=1}^{3}
\left\{ \phi_{\vec{n}+\vec{e_{i}}} + \phi_{\vec{n}- \vec{e}_{i}}
- 2 \phi_{\vec{n}} \right\} - m^2 \phi_{\vec{n}}, \label{1.1} \\
\mbox{with} \; \vec{n} & = & \left( n_1,n_2,n_3 \right) \; \;
, n_i \in \mbox{\bf Z}, \nonumber
\end{eqnarray}

where $\phi_{\vec{n}}(t) = \phi (a \vec{n},t)$ is a function defined
on lattice points only.
Now we imbed the lattice in a continuous space. Then we can
replace differences by the operation
\beq
 \frac{f(x+a)-f(x-a)}{2a} = \frac{1}{a} \sinh (a \partial_{x} ) f(x).
\label{1.2}
\eq
In this way the difference equation is transformed into
\beq
\left\{\pl_t^2 - \left(\frac{2v}{a}\right)^2 \sum_{j=1}^3 \sinh^2\left(
\frac{a}{2}\, \pl_{x_j}\right) + m^2 \right\} \phi (\vec{x},t)
= 0 , \label{1.3}
\eq
where $ v= \omega a $. In the limit $a \rightarrow 0$ we find the Klein--Gordon
equation. (On the other hand any solution of eq.(\ref{1.3}) satisfies
eq.(\ref{1.1}) at the lattice points.)
The operator of the Klein--Gordon equation is a Casimir operator
of the Poincar\'e  algebra. Consequently  scalar fields belonging to an
irreducible representation of the Poincar\'e algebra must be solutions of this
partial differential equation \cite{sex}.
We claim to find the same situation in the lattice case.

With exception of the inhomogeneous part, the generators
of the deformed Poincar\'e algebra are constructed in the same way as
the deformed Klein--Gordon equation.

Their representation as differential operators is:
\begin{eqnarray}
p_0 & = &  \frac{i}{v}\, \pl_t ,\nonumber \\
k_j & = &  \exp \left( i\, a \, p_j \right) , \mbox{with } p_j = -i\pl_{x_j},
\nonumber \\
K_j & = &  i \left( \frac{x_j}{v} \, \pl_t + \frac{vt}{a} \sinh (a\pl_{x_j})
\right), \nonumber \\
J_i & = &  -i \left( \sum_{jk} \epsilon_{ijk} \frac{x_j}{a} \sinh (a\pl_{x_k})
\right), \label{gl.4}
\end{eqnarray}
where indices run from 1 to 3.
The deformed Poincar\'e algebra is generated by the elements
$ {\bf 1}, p_0, k_i, k_i^{-1}, K_i, J_i$. The momentum generator in the
continuous limit is not reproduced by $\lim_{a\to\ 0} k_j$ but rather with
$\lim_{a\to\ 0} \:i/a \: (1-k_j)$.
To calculate the commutation relations it is useful to know that
\beq
\left[ e^{a\pl_x} , x \right] = a e^{a\pl_x} .
\eq
This gives
\begin{eqnarray}
\left[\cosh (a\pl_x), x \right] & = &  a \sinh (a\pl_x), \nonumber \\
\left[\sinh (a\pl_x), x \right] & = &  a \cosh (a\pl_x). \nonumber
\end{eqnarray}
Using this we find
\begin{eqnarray}
\left[ p_0, k_i \right] & = &  0 , \hspace{1cm} \left[ k_i, k_j \right] \;
= \;0,\nonumber \\
\left[ k_i, K_j \right] & = &  \delta_{ij}\: a \: p_0\, k_i ,\nonumber \\
\left[ K_i, K_j \right]  & = &  \left( \frac{x_i}{a} \sinh (a\pl_{x_j}) -
\frac{x_j}{a} \sinh (a\pl_{x_i}) \right) \nonumber \\
               & = &  -i \sum_k \epsilon_{ijk} J_k , \nonumber \\
\left[ p_0, K_i \right] & = &  \frac{1}{2a} \left( k_i^{-1} - k_i \right),
\nonumber \\
\left[ J_1, J_2 \right] & = &  - \left( \frac{y}{a} \cosh (a\pl_z)
\sinh (a\pl_x) - \frac{x}{a} \cosh (a\pl_z) \sinh (a\pl_y)\right) \nonumber \\
               & = &  \frac{i}{2} (k_3 + k_3^{-1}) J_3, \hspace{1cm}
                      \mbox{and cycl. perm. of the indices } , \nonumber \\
\left[ K_1, J_2 \right] & = &  - \left[ \frac{vt}{a} \sinh (a\pl_x),
\frac{x}{a} \sinh (a\pl_z) \right] + \left[ \frac{x}{v} \, \pl_t,
\frac{z}{a} \sinh (a\pl_x) \right]  \nonumber \\
               & = &  - \frac{vt}{a} \cosh (a\pl_x) \sinh (a\pl_z) -
                        \frac{z}{v} \cosh (a\pl_x) \pl_t \nonumber \\
               & = &  \frac{i}{2} (k_1 + k_1^{-1}) K_3, \hspace{1cm}
\mbox{and cycl. perm. of the indices } , \nonumber \\
\left[ k_i, J_j \right] & = &  \frac{i}{2} \epsilon_{ijn} k_i (k_n - k_n^{-1}).
\end{eqnarray}
We have sometimes replaced $x_i$ by $x,y,z$ to avoid indices.
The Jacobi identities follow by associativity of operator muliplication.
Now we check whether the length of a modified 4-momentum still provides a
Casimir operator. We define
\beq
\tilde{p}_{\mu}   =    \left\{ \begin{array}{l} -ip_0 \nonumber \\
2i/a\: \sin
 \left( a/2 \: p_j \right),\hspace{0.5cm} j= 1,2,3  \hspace{0.5cm},
\end{array} \right.
\eq
and get
\begin{eqnarray}
C & = &  \tilde{p}^{\mu} \tilde{p}_{\mu} \nonumber \\
  & = &  p_0^2 - \left( \frac{2}{a} \right)^2 \sum_j \sin^2\left(\frac{a}{2}
         \, p_j \right) \nonumber \\
  & = &  p_0^2 - \left( \frac{1}{a} \right)^2 \sum_j \left(k_j + k_j^{-1} -2
\right) ,
\end{eqnarray}
where we used the metric $\eta_{\mu \nu} = \mbox{ diag}(-1,1,1,1)$.
It is sufficient to check its commutators with
the boosts and the angular momentum generators. Taking for example $K_1$
we find
\begin{eqnarray}
[ C, K_1] & = &  2 p_0 [ p_0, K_1 ] + \frac{1}{a^2} \sum_j [ k_j + k_j^{-1} -2,
                 K_1 ] \nonumber \\
          & = &  \frac{1}{a} \, p_0 ( k_1^{-1} - k_1 ) + \frac{1}{a} (k_1 p_0 -
                 k_1^{-1} p_0 ) \nonumber \\
          & = &  0.
\end{eqnarray}
Commuting $C$ with an angular momentum generator gives
\begin{eqnarray}
[C, J_1 ] & = &  [ p_0^2, J_1 ] + \frac{1}{a^2} \sum_j [ k_j + k_j^{-1} -2,
                 J_1 ] \nonumber \\
          & = &  \frac{i}{a^2} \{ - k_2 (k_3 - k_3^{-1}) + k_3 (k_2 - k_2^{-1})
                 \nonumber \\
          &   &  + k_2^{-1} (k_3 - k_3^{-1}) - k_3^{-1} (k_2- k_2^{-1}) \}
                 \nonumber \\
          & = &  0 ,
\end{eqnarray}
which proves that $C$ is indeed a Casimir operator.

\parskip3ex
The real structure, i.e. the existence of an antilinear
antimultiplicative involution,
is inherited from the Heisenberg algebra in four dimensions.
In the infinite dimensional representation given in (\ref{gl.4})
the angular momentum as well as the boost generators are shown to be hermitian
while for the translation operators one finds
\beq
      k_i^* = k_i^{-1}.
\eq
Thus $C$ is hermitian as well.

Consistency of this $*$ - structure is found by checking the algebra relations:
\begin{eqnarray}
\left[ k_i, K_j \right]^*  & = &  -\left[ k_i^{-1}, K_j \right] = \delta_{ij}
                                  \: a \: p_0\, k_i^{-1} \nonumber \\
                           & = &  (\delta_{ij}\: a \: p_0\, k_i)^* ,\nonumber
\\
                    \left[ K_i, K_j \right]^*  & = &  -\left[ K_i, K_j \right]
          =  \left( -i \sum_k \epsilon_{ijk} J_k \right)^*, \nonumber \\
\left[ p_0, K_i \right]^* & = &  \frac{1}{2a} \left( k_i - k_i^{-1} \right) =
                      \frac{1}{2a} \left( k_i^{-1} - k_i \right)^* \nonumber \\
\left[ J_1, J_2 \right]^* & = &  - \left[ J_1, J_2 \right] \nonumber \\
                          & = & \left( \frac{i}{2} (k_3 + k_3^{-1}) J_3
 \right)^*, \hspace{1cm} \mbox{and cycl. perm. of the indices } , \nonumber \\
\left[ K_1, J_2 \right]^* & = & - \frac{i}{2} (k_1 + k_1^{-1}) K_3  \nonumber
\\
                          & = &  \left( \frac{i}{2} (k_1 + k_1^{-1}) K_3
   \right)^*, \hspace{1cm} \mbox{and cycl. perm. of the indices } , \nonumber
\\
\left[ k_i, J_j \right]^* & = & - \left[ k_i^{-1}, J_j \right] = \frac{i}{2}
                           \epsilon_{ijn} k_i^{-1} (k_n - k_n^{-1}) \nonumber
\\
                          & = & \left(- \frac{i}{2} \epsilon_{ijn} k_i^{-1}
                              (k_n - k_n^{-1}) \right)^*.
\end{eqnarray}

\parskip1.5ex
{\bf Conclusion} We have found a deformation of the Poincar\'e algebra with
the deformation parameter given by the lattice spacing.
Of course this is not its quantum deformation since a coproduct and
antipode allowing the continuous limit are still missing.

\parskip2ex
{\bf \large Acknowledgements: }

I would like to thank Julius Wess for drawing my attention to this problem and
Ralf Weixler and Werner Zippold for helpful discussions.

\end{document}